\documentclass[floatfix,twocolumn,preprint,aps]{revtex4}
\usepackage{graphicx}
\usepackage{psfrag}

\begin{document}
\begin{abstract}
We use a new version of the reversible Parking Lot Model to study the compaction of
vibrated polydisperse media. The particle 
sizes are distributed according to a truncated power law. We introduce a
self-consistent desorption mechanism with a hierarchical initialization of
the system. In this way, we approach densities close to unity. The final density
depends on the polydispersity of the system as well as on the initialization and  
will reach a maximum value for a certain exponent in the power
law.
\end{abstract} 


\title{The reversible polydisperse Parking Lot Model}

\author{Martin Wackenhut, Hans Herrmann}
\affiliation{Institute for Computational Physics, University of
Stuttgart,    
Pfaffenwaldring 27, 70569 Stuttgart, Germany}

%  \and \\ICA1, Universität Stuttgart\\
%Pfaffenwaldring 27, 70569 Stuttgart, Germany\\
%M.Wackenhut@ica1.uni-stuttgart.de\\
%H.J.Herrmann@ica1.uni-stuttgart.de}

\date{\today}

\maketitle

\section{Introduction}
	The vibratory compaction of granular materials has long been of importance
	in technological applications like high performance concrete or ceramics
	which have to withstand extreme stress.  Extensive experimental studies
	have been conducted on monodisperse systems.  Knight \textit{et al.}
	\cite{Physrev_E_51_5_3957} investigated density relaxation of a column
	containing monodisperse spherical beads subject to a long sequence of taps
	while Nowak \textit{et. al} \cite{Physrev_E_57_2_1971}  studied density
	fluctuations in vibrated granular materials.
	
	Several models were introduced to describe the dynamics of a granular
	system under compaction, including a lattice model
	\cite{PhysRev_E_59_6_6830} and the Tetris model
	\cite{PhysRevLett_79_8_1575} as a special case of a frustrated lattice
	gas.
	The study of polydisperse systems reaches back to the ancient Greeks where
	Apollonius of Perga studied the problem know as 'Apollonian
	packing' \cite{Aste,Powder_Tech_102_274}. This problem deals with the question how to tile the
	space with circles by iteratively placing, between every three circles, a
	circle tangentially touching all three.  The Apollonian packing is a
	special case of the so called 'space filling bearings' (SFB) in which a plane
	is tiled with circles touching one another such that the whole area is
	covered with circles \cite{PhysRevLett_65_26_3223,
	JPhys_A_33_1417, JPhys_A_24_L481, Frac_2_4_521}.  These space filling
	bearings fill space with a particle	size distribution given by a truncated power law.
	Different studies of polydisperse packings where
	conducted by Aste \cite{PhysRev_E_53_3_2571}, Dodds and Weitz \cite{PhysRev_E_65_5_56108} and Brilliantov \textit{et al.}
	\cite{PhysRevLett_76_21_4058,PhysRev_E_58_3_3530}. 

	The aim of our work is to study the time evolution of density in
	polydisperse systems under vibratory compaction. The size distribution of
	the particles obeys a truncated power law (different than the one used
	by Brilliantov).
	We modified the one dimensional reversible Parking Lot Model
	\cite{JChemPhys_100_9_6778, PublMathResInstHungAcadSci_3_109,
	PublMathResInstHungAcadSci_5_353} where identical particles adsorb on an
	interval with an adsorption rate $k^+$ and desorb with a rate $k^-$, such
	that the system is hierarchically initialized and the equilibrium state
	with the highest density is reached through self-consistent desorption.

\section{The Model} \label{sec:model}

	We first explain the classical Parking Lot Model \cite{PublMathResInstHungAcadSci_3_109} 
	and its reversible variant \cite{JChemPhys_100_9_6778}.
	Afterwards we introduce our generalization to polydisperse systems 
	and the self consistent desorption probability.
	
	\subsection{The Parking Lot Model}
		\psfrag{A}{$l$}
		\psfrag{k+}{$k^+$}
		\psfrag{k-}{$k^-$}

		In the classical Parking Lot Model (PLM), identical particles of size $r$
		try to adsorb with an adsorption rate $k^+$ on randomly chosen places along an
		interval of length $l$. The adsorption fails if the chosen place is
		partially occupied by a previously adsorbed particle.
		Because of the irreversible adsorption mechanism the system will reach a
		so called jammed state with a final density of $\rho_{jam}\approx
		0.7475$.

\begin{figure}[h!] 
			\centering 
			\includegraphics[width=8. cm]{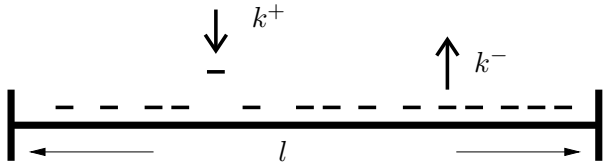}
			\caption{The reversible Parking Lot Model with an interval length $l$
							and particles of equal size $r$. Particles adsorb with a rate $k^+$ 
							and desorb with a rate $k^-$.}
			\label{fig:RPLM-xfig}
		\end{figure}

		Krapivsky and Ben-Naim extended this model to the reversible Parking Lot Model
		\cite{JChemPhys_100_9_6778}. Here particles will additionally 
		desorb with a desorption rate $k^-$ (Figure \ref{fig:RPLM-xfig}) and the
		density of the system's equilibrium state can be different from
		$\rho_{jam}$. This equilibrium state is determined by the ratio between
		the adsorption and the desorption rates $k=k^+/k^-$.

		\noindent
   	For the final density $\rho_{ss}$ the following leading behavior in the two
   	limiting cases was found:
     \begin{equation}
       \rho_{ss} (k)\cong \left\{ \begin{array}{ll} 
       					  k & \textrm{for } k \ll 1 \\  
       					  1-ln(k)^{-1} & \textrm{for } k \gg 1 
       			  \end{array} \right .
           \label{rho_PLM} 
     \end{equation}
		In the limit of $k \rightarrow \infty$ ($k^+=1, k^- \rightarrow 0$),
		also called the desorption controlled limit, the interval is
		completely filled with particles and $\rho_{ss}=1$ when time goes to
		infinity. 
	
	\subsection{Generalization to a polydisperse Model}
	 	Granular materials that reach highest densities consist of particles
		with different sizes. Such media are called polydisperse. 
		We will simulate such media using a polydisperse PLM.
		
\begin{figure}[h!] 
			\centering 
			\includegraphics[width=8. cm]{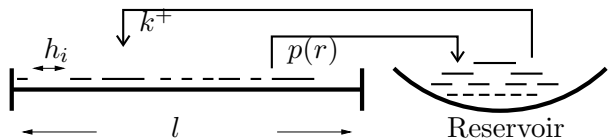}
			\caption{The self consistent reversible polydisperse Parking Lot Model with an interval
							length $l$, particles of different size $r$ and the reservoir. 
							Particles adsorb with a rate $k^+$ and desorb with a conditional
							probability $p(r)$ which also depends on the distribution of holes in the
							interval.}
			\label{fig:self-xfig}
		\end{figure}

		The behavior of a polydisperse PLM is governed by the size distribution
		of the particles, the amount of particles available in the system, the
		initialization process and the dependence of the desorption rate on the
		particle size.
		Figure \ref{fig:self-xfig} shows the setup for the polydisperse Parking
		Lot Model.
		
		\psfrag{k-}{$p(r)$}
		\psfrag{Reservoir}{Reservoir}
		\psfrag{hi}{$h_i$}

			Unlike to the case of monodisperse systems we introduce a reservoir because 
			it turns out to be necessary to restrict the number of	particles of each size in
			order to impose a given size distribution. 
			For a system with $K$ particles of which
			$K_1$ are adsorbed on the interval, this reservoir will contain the
			remaining $K_2=K-K_1$ particles.
			
			We restrict the size of the particles to the interval 
			$\overline{R} =	[r_{min},r_{max}]$ which we will refer to as
			the 'range of particle sizes'. For all simulations
			discussed here,	$r_{max}$ is set to unity and
			the size distribution of all particles is given by
  		a truncated power law: 
      \begin{equation}
	      P(r)=a\cdot r^{-b}, \qquad r \in \overline{R}
	      \label{eq:powerlaw}
	    \end{equation}
   		Where $a$ is a normalization factor chosen such that $\int_{r_{min}}^{r_{max}}P(r)dr=l$. We allow the exponent $b$ to
   		take values in the range of $1.306 < b < 1.802$, the same range as
   		found for two dimensional space filling bearings \cite{JPhys_A_33_1417}.
			We also apply periodic boundary conditions to the system. Its density $\rho$
			is given by.  
			\begin{equation}\rho=\sum_{i=1}^{K_1} r_{i}/l \end{equation}
			where $K_1$ is the total number of adsorbed particles and $r_i$ is the
			length of the $i$th adsorbed particle.
			
			As in the classical PLM, the particles attempt to be adsorbed with the
			adsorption rate $k^+$. This rate is defined as the number of particles
			attempting to be adsorbed per time unit.
			A particle of given size $r$, randomly chosen among the adsorbed
			particles, will desorb with a conditional desorption probability $p(r)$.
			This conditional probability can be calculated through:
			\begin{equation}p(r)=\sum_{i=1}^{K_1}{'}(h_i - r)/l \label{eq:prob}\end{equation}
   		Here $K_1$ is the total number of holes which is the same as the number
			of adsorbed particles when using periodic boundary conditions. The
			variable $h_i$ is the length of a single hole, which can be zero 
			in the case of two touching particles.
			%The value of $j$ in the sum is the number of holes with a size $h_i$. 
   		The primed sum only considers terms with $h_i>r$.  Thus $p(r)$ will be
   		zero if the particle size $r$ is larger than any available hole.  We
   		call $p(r)$ the 'self consistent desorption probability' because of
   		its dependence on the hole distribution. It changes in time as the
   		configuration in the interval changes.

			This approach is justified because in an experiment, a particle
			can only leave its place if there is a large enough hole where it can
			move to. The holes of the system are stored in a list $H$ which is
			updated after each adsorption and desorption.
			
			In the monodisperse Parking Lot Model a constant desorption rate
			removes particles regardless of their local environment. Therefore,
			the system ultimately reaches an equilibrium state independent of the
			initial condition.  This is not the case in a polydisperse system
			where the self consistent desorption is implemented.  If we would use
			a constant desorption rate in the polydisperse system, the result
			would be a final density lower than the $\rho_{jam}$ obtained in the
			monodisperse model.

		\subsubsection{Dynamics}			
			Now we will explain the dynamics of the 
			system. The word 'random' will denote an equally distributed
   		random number generator unless stated otherwise.

			First the particles available to the system are generated.
			For this a total number of $K$ particles is put into
			an initially empty reservoir (list). The size $r_k$ of each particle 
			is sampled randomly from a power
			law distribution Eq. (\ref{eq:powerlaw}).  The total number $K$ of
			particles is given by $\sum_{k=0}^Kr_k=l$.  Thus the generation of
			particles is stopped as soon as the sum of the lengths of all particles
			is equal to $l$. 
			We call the size distribution of particles in the reservoir at this
			time ($t=0$) the '$t_0$ distribution'.
			
  		Next an initial distribution of particles on the interval is generated. 
			We will call this the 'initialization of the interval'.  This
			is necessary because for an almost empty interval $p(r)$ is close to
			unity for all particles and any adsorbed particle would most likely
			desorb again. A further justification for this
			initialization is given in section \ref{sec:sizedist}.
						
			In order to reach highest densities a hierarchical initialization is
  		used. This idea was inspired during a visit to W. Losert's \cite{losert}
  		laboratory at the University of Maryland. The reservoir
  		is sorted by size such that $r_1 > r_2 > \cdots > r_K$. Then each
  		particle, starting with the biggest particle ($r_1$), is given $I$
  		trials to adsorb. During the initialization, the desorption
  		probability is set to zero. This resembles the filling of a recipient
  		in an experiment, where the large particles are loaded first and the
  		system is not vibrated.
   		
   		For each adsorption trial, a random point in the interval is chosen. If
      the chosen place is not even partially occupied the particle will adsorb.
      If the particle did not adsorb after its
      $I$th trial it is left in the reservoir and one continues with the
      next smaller one. The initialization is finished when each
      particle had up to
      $I$ trials for adsorption. The density resulting from this
      initialization is called $\rho_{init}$.
      
			Now the adsorption/desorption mechanism is activated.  With a rate
			$k^+$, a particle is randomly picked from the reservoir and adsorbed
			if possible on a random position. If the adsorption is not possible,
			the particle does not get a second trial but will be put back into the
			reservoir.
      For desorption a particle on the interval is randomly picked at each
      time step.  With the probability $p(r)$, given by Eq. (\ref{eq:prob}), it is put back into the
      reservoir.  Each time the distribution of particles in the interval
      changes the list of holes $H$ is updated accordingly. 
      
      In this model we have two different definitions of time. 
      The first definition covers the initialization of the system. Here one time step
      consists of a single adsorption trial. The second definition
      considers the time after the initialization (the dynamics).
      Here a single time step in the system consists of an adsorption
      and a desorption trial. Adsorption is attempted with the rate $k^+$
      while desorption is checked every time step.
      These two definitions are of significant difference. Nevertheless
      we will put them on the same scale in our
			density plots in order to investigate dependencies on different
			parameters.
      
\section{Monte Carlo results} \label{sec:mc1d}
  	In this section we give an overview over the performed simulations.
		We always use an adsorption rate
		$k^+=1$. Thus, one adsorption trial is performed during each time step.

		\psfrag{l}{$l$}
		\psfrag{time}{time}
		\begin{figure}[h!] 
			\centering 
			\includegraphics[width=8. cm]{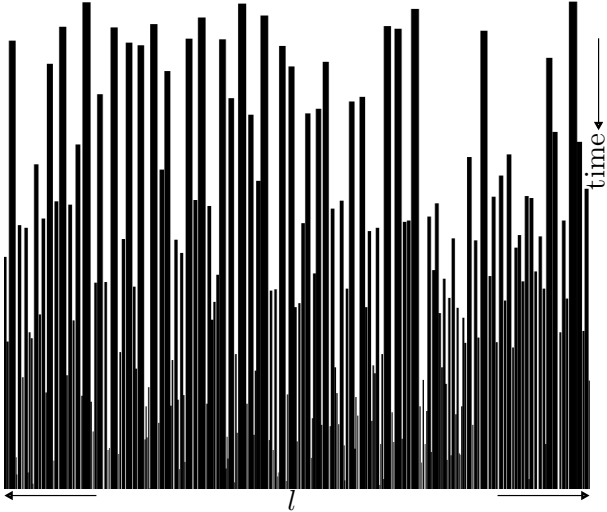}
			\caption{Spatio-temporal diagram of the
			  			initialization process.  System parameters: $l=300$,
  						$r_{min}=0.001$, $r_{max}=1$, $k^+=1$, $I=1000$, $b=1.33$, $K=5420$. 
			  			Density at the end of the initialization: $\rho \approx
							0.985$}
			\label{fig:system}
		\end{figure}

    The alignment of the particles on the interval $l$ at a certain time is
    visualized in a spatio-temporal diagram by plotting at each time step
		the system by a height of one pixel and a width
    proportional to $l$. The density is gray scale encoded using white
    pixels for zero density and black pixels for density one.  The
    darkness of the pixel depends linearly on the density.  Arranging
    these pictures in chronological order results in the spatio-temporal
    diagram as displayed in Figure \ref{fig:system} which shows the time
    evolution of the system during the initialization
    process.

		Because of the hierarchical initialization the big particles, shown as
		regions of high density, are adsorbed first.  As time
		progresses the interval is consecutively filled with smaller and smaller
		particles which settle in the remaining gaps until the system reaches a
		high density (desorption is always turned off in this case).

	\subsection{Size Distribution} \label{sec:sizedist}
		\begin{figure}[h!] 
			\centering 
			\includegraphics[width=8. cm]{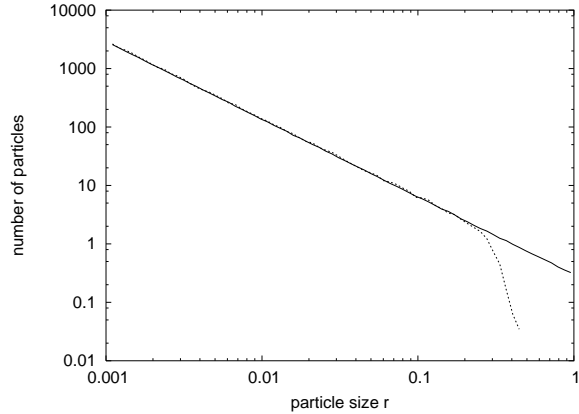}
			\caption{Particle size distribution for the self consistent system (solid
							line, $I=1000$) and a system with a constant desorption rate using
							no initialization (dotted line). In both cases we display the
							distribution at the end of the simulation.}
			\label{fig:probability}
		\end{figure}

  	The hierarchical initialization avoids the exclusion of large
		particles and thus leads to an adsorption of almost
		all particles.
    In the ideal case of a completely filled interval, the $t_0$
    distribution and the distribution of adsorbed particles for a system
    in its final state would be identical.
  	
  	In Figure \ref{fig:probability} we display the size distribution of the
  	adsorbed particles at the end of the simulation. The solid line shows
  	the distribution for a self-consistent desorption rate as given by
  	Eq. (\ref{eq:prob})
  	at the end of the simulation which coincides very well
  	with the $t_0$ distribution which is omitted for clarity. All
  	simulations show the same overall behavior for different values of
  	$\overline{R}$ and $b$.  In addition we show the distribution for a
  	system with constant desorption rate using no initialization. For large
  	particles the distribution deviates very strongly from the $t_0$
  	distribution which justifies our hierarchical initialization.

	\subsection{Desorption Probability}
		\begin{figure}[h!] 
			\centering 
			\includegraphics[width=8. cm]{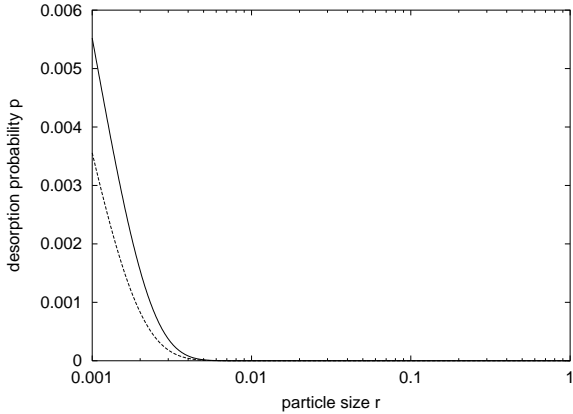}
			\caption{Desorption probability of the self consistent system as a function
							of the particle size in a semi-logarithmic plot.  For a denser system the probability is
							smaller.  The solid line is the probability at the end of the
							initialization process ($t\approx 10^7$). 
							The dotted line shows $p(r)$ at the end of
							the simulation ($t\approx10^9$). System parameters: 
							$l=5000$, $r_{min}=0.001$, $r_{max}=1$, $k^+=1$, $I=1000$, $b=1.33$,
  						average $K=89618$.
			 				Data are averaged over 10 runs.}
			\label{fig:desprob}
		\end{figure}
    The desorption probability $p(r)$ right after the initialization of the
    interval and at the end of the simulation is shown in Figure \ref{fig:desprob}.
		The probability depends on the hole distribution and
		therefore changes in time.  The solid line in Figure \ref{fig:desprob} shows the
		probability at the end of the initialization process ($t\approx 10^7$).
		At this time, the transition from the initialization to the
		adsorption/desorption process takes place.
		The dotted line shows $p(r)$ at the end of our simulation
		($t\approx 10^9$). The probability $p(r)$ decreases the denser the system gets.

  	The probability $p(r)$ will become zero for large particles 
  	at some time and they will not be able to desorb
  	anymore. This leads to a stable configuration with highest densities
  	including all the large particles in the interval.

	\subsection{Density evolution}
		\begin{figure}[h!] 
			\centering 
			\includegraphics[width=8. cm]{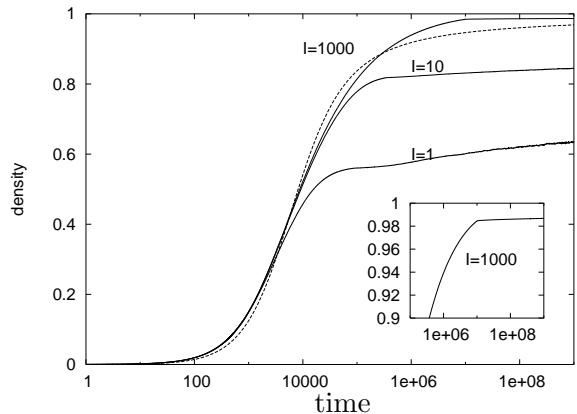}
			\caption{Density as a function of time in the self-consistent system for
			  			different values of $I$. The dotted line is the fitted curve according to
  						Eq. (\ref{eq:evolution}).  Fit parameters:	$\rho_\infty=1$, $\Delta
			  			\rho_\infty = 0.967$, $B=2.829$, $\tau=19339$.  
  						Data are averaged over 10	runs.
			  			The system parameters are the same as in Figure \ref{fig:desprob}.}
			\label{fig:density}
		\end{figure}

    The development of the density is shown in 
    Figure \ref{fig:density}. Here the density evolution for different
    numbers of trials $I$ are shown. The inset enlarges the time interval
		where the transition from the initialization to the
		adsorption/desorption process takes place.  From this point on,
		adsorption- and desorption take place simultaneously; the density
		rises slower.
		
  	Knight \cite{Physrev_E_51_5_3957} and Nowak \cite{Physrev_E_57_2_1971}
  	measured experimentally that the time evolution of density in
  	monodisperse systems can be asymptotically fitted by the expression
 		\begin{equation}\rho(t)=\rho _\infty-\frac{\Delta\rho_\infty}{1+Bln(1+t/\tau)} \label{eq:evolution} \end{equation}
  	where the parameters $\rho_\infty$, $\Delta \rho_\infty$,  $B$ and $\tau$
		depend on the experimental setup.  Here $\rho_\infty$ is the final density while $\Delta
		\rho_\infty$ is the difference between this value and the density at $t=0$. The parameters $B$ and
		$\tau$ fit the logarithmic behavior of the curve.
		The same asymptotic behavior was obtained by Krapivsky and
  	Ben-Naim for the reversible parking lot model using a variable desorption
  	constant \cite{Physrev_E_57_2_1971}. 
  	The dotted line in Figure \ref{fig:density} is the fit using Eq.
		(\ref{eq:evolution}).
		We can see that the density after the initialization is very close to the
		final density. Thus the dominating part in our simulations is the
		initialization process which we will investigate closer in the following
		section.

	\subsection{Discussion} \label{sec:discussion}
		Using the self consistent desorption, the final density mostly depends
		on the initialization. Large particles that could not adsorb during the
		initialization are, in general, excluded from the interval for the rest
		of the simulation. The initialization is characterized by $\rho_{init}$.
		Three parameters determine this density. The number of trials $I$, the
		exponent in the power law $b$ and the range of particle sizes
		$\overline{R}$.
		
		\subsubsection{Dependence on $I$}
		\begin{figure}[h!] 
			\centering 
			\includegraphics[width=8. cm]{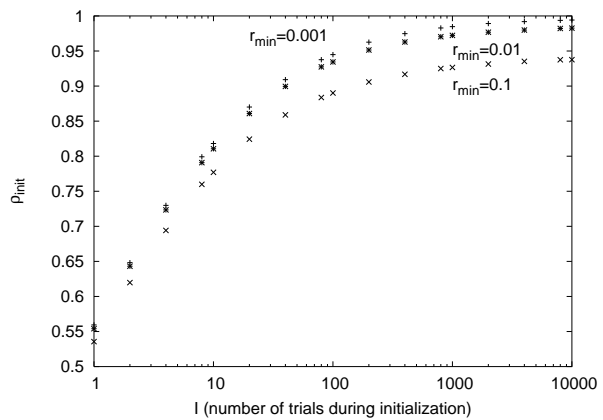}
			\caption{Dependence of $\rho_{init}$ on the number of trials $I$.
							Displayed are three curves with different $r_{min}$ 
							while $r_{max}=1$ is kept constant for all three runs. The maximal
							possible density 
							is determined by $r_{min}$ and reached for very large $I$.
							System parameters: $l=5000$, $r_{max}=1$, $k^+=1$,
  						$b=1.33$ and $K\approx89000$. Data are averaged over 9
							independent runs.}
			\label{fig:I_ana}
		\end{figure}

\begin{figure}[h!] 
			\centering 
			\includegraphics[width=8. cm]{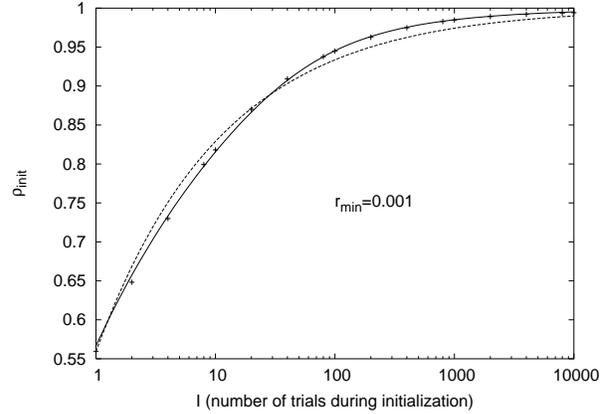}
			\caption{Simulation data of Figure \ref{fig:I_ana} fitted by 
							Eq. (\ref{eq:fit1})	(dotted line,	$\rho_{max}=1$, $\Delta \rho=0.4406$ , $f=0.4109$)
							and Eq. (\ref{eq:fit2}) (solid line, $\rho_{max}=1$, $\Delta \rho=0.4406$, $B=0.636$, $\tau=21.3$, $f_n=0.3163$.) 
							The system parameters are the same as in Figure \ref{fig:I_ana} }
			\label{fig:I_ana_fit}
		\end{figure}

			As already seen in Figure \ref{fig:density} a larger number of trials $I$ 
			increases considerably the final density $\rho _{init}$ of the system.
			In Figure \ref{fig:I_ana} the dependence of $\rho_{init}$ on $I$ is displayed. 
			Starting from a low density of about $0.55$ the density of
			the system increases until it reaches its maximum
			density for this value of $r_{min}$. 
			Because the system is hierarchically initialized the final density 
			depends on $I$.
			This dependence can be fitted by:
			\begin{equation}\rho_{init}(I)=\rho_{max}(r_{min})-\Delta \rho \cdot I^{-f} \label{eq:fit1} \end{equation}
			Here $\rho_{max}(r_{min})$ is the maximum density the system can reach for the
			chosen value of $r_{min}$ in the limit $I\rightarrow \infty$ and $\Delta
			\rho$ is the difference between $\rho_{max}(r_{min})$ and
			the initialization density for $I=1$.
			The exponent $f$ is a fit parameter ($f=0.4406$ for $r_{min}=0.001$) giving the logarithmic relaxation
			towards the highest density.
			In Figure \ref{fig:I_ana_fit} the curve fitted with Eq.
			(\ref{eq:fit1}) is the dotted line.

			A better fit can be accomplished when in Eq. (\ref{eq:fit1}) the last
			term is multiplied by the last term given in Eq. (\ref{eq:evolution}). The
			resulting expression has the form:
			\begin{equation} \rho_{init}(I)=\rho_{max}-\frac{\Delta \rho}{1+B\cdot ln(1+I/\tau)}I^{-f_n} \label{eq:fit2} \end{equation}
			In Figure \ref{fig:I_ana_fit} this results in the full line coinciding with the
			simulation data.
			The value of $f_n$ is $f_n=0.3163$ for $r_{min}=0.001$.

		\begin{figure}[h!] 
			\centering 
			\includegraphics[width=8. cm]{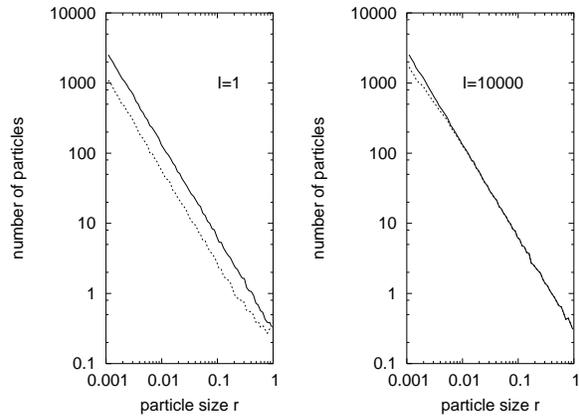}
			\caption{Comparison of the size distribution at the end of the
							initialization for a different number of trials $I$
							(dotted line) with the $t_0$ distribution (solid line).
							System parameters: $l=5000$, $r_{min}=0.001$, $r_{max}=1$, $k^+=1$,
  						$b=1.33$ and $K\approx89000$.}
			\label{fig:I_ana_hist}
		\end{figure}	
			
			The number of trials $I$ also influences the size distribution of the
      particles. In Figure \ref{fig:I_ana_hist} the size distribution for two
      different values of $I$ is displayed. The distribution of adsorbed
      particles at the end of the simulation coincides better with the $t_0$
      distribution for larger $I$.

      The deviation from the original distribution depends on the value of $I$
      because a larger $I$ gives each particle more trials for
			adsorption. For large $I$,  big particles adsorb first because of the
      hierarchical initialization and a deviation is only
      noticeable for small particles. Comparing Figure \ref{fig:I_ana_hist} with Figure \ref{fig:probability} reveals that
			the deviation from the $t_0$ distribution (large $I$) after the
			initialization almost vanishes when the adsorption/desorption
			mechanism is activated.

			Finally we study which effect $I$ has on the self consistent
			desorption probability $p(r)$.
		\begin{figure}[h!] 
			\centering 
			\includegraphics[width=8. cm]{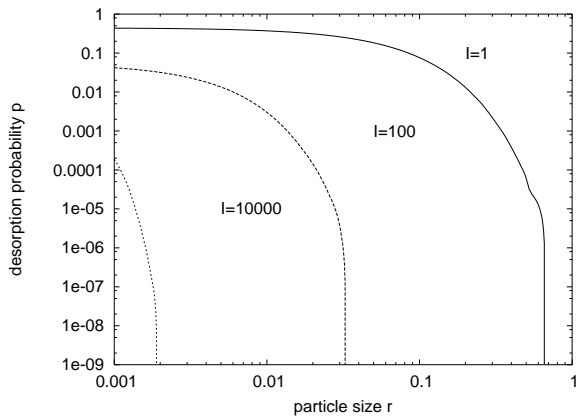}
			\caption{Desorption probability for different number of trials $I$ in a
					double logarithmic plot, right
					after the initialization.  It is zero for the largest particles in
					all three cases.  System parameters: $l=5000$, $r_{min}=0.001$,
					$r_{max}=1$, $k^+=1$,
  				$b=1.33$, and $K\approx89000$. }
			\label{fig:I_ana_des}
		\end{figure}	
				
      In Figure \ref{fig:I_ana_des} the desorption probability right after the
			initialization is displayed for three different $I$.  The larger $I$
			the smaller the overall desorption probability.
			More trials $I$ during
			the initialization result in smaller holes and a denser system.  Hence
			the overall probability will be lower for a larger $I$. In all three
			runs it is zero for the largest particles.
			
	\subsubsection{Dependence on $\overline{R}$} \label{sec:R}
		\begin{figure}[h!] 
			\centering 
			\includegraphics[width=8. cm]{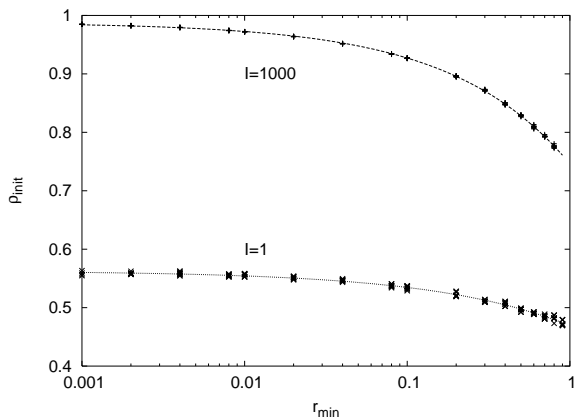}
			\caption{Dependence of $\rho_{init}$ on $r_{min}$ fitted by Eq. (\ref{eq:fit3})
							($\rho_{max}(I)=0.987$, $\Delta	\rho=0.241$, $e=0.600$ for $I=1000$ and
							$\rho_{max}(I)=0.562$, $\Delta \rho=0.093$, $e=0.519$ for $I=1$).
							The smallest particle size
							$r_{min}$ is varied while $r_{max}$ is kept constant.
							System parameters: $l=5000$, $r_{max}=1$, $k^+=1$,
  						$b=1.33$, $K\approx5200 - 90000$. Data are averaged over 9
							runs.}
			\label{fig:r_min_ana-fit}
		\end{figure}	
	
			The maximum density also depends on the range of particle sizes
			$\overline{R}$.  In Figure \ref{fig:r_min_ana-fit} $r_{max}$ was kept
			constant (unity) while $r_{min}$ was varied.  The dependence of $\rho_{init}$
			on $r_{min}$ is displayed for two different values of $I$.  A larger
			range of particle sizes $\overline{R}$ results in a higher
			$\rho_{init}$. This general behavior is independent of the 
			number of trials $I$.
			
      The hierarchical initialization, starting with the biggest particle,
      allows the system to densify better because the holes get smaller and
      smaller. Would the initialization start with the smallest particles
      first, the system would end up in a final state with an
      extremely low density.
      The smallest size $r_{min}$ in the range of particle sizes $\overline R$
			determines the smallest hole that can be filled and thus determines
			the highest reachable density.
			
      The dependence on the minimal particle size in
      Figure \ref{fig:r_min_ana-fit} can be fitted very well by the equation.
			\begin{equation}\rho_{init}(r_{min})=\rho_{max}(I)-\Delta \rho \cdot r_{min}^{e} \label{eq:fit3} \end{equation}
			Here $\rho_{max}(I)$ is the maximal possible density for the applied
			number of trials $I$ which occurs for $r_{min}\rightarrow 0$. The
			variable $\Delta \rho$ is the difference between this maximal density
			and the density the monodisperse system reaches for the same applied number
			of trials $I$. Thus we can determine these values through averaging over
			several independent simulations. The fit parameter $e$ has a value of
			$e=0.6$ for $I=1$ and
			$e=0.519$ for $I=1000$.

		\subsubsection{Dependence on $b$}
		\begin{figure}[h!] 
			\centering 
			\includegraphics[width=8. cm]{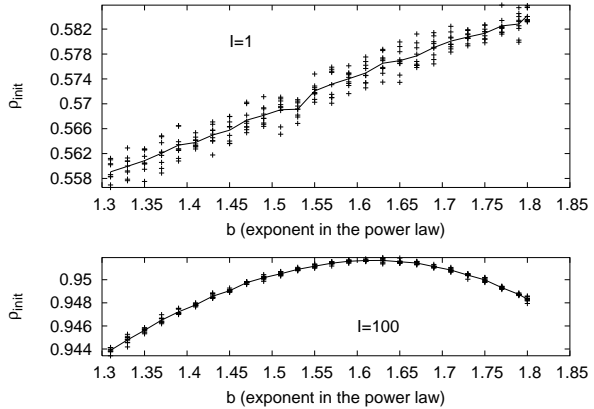}
			\caption{Dependence of $\rho_{init}$ on the exponent $b$.
      					System parameters: $l=5000$, $r_{min}=0.001$, $r_{max}=1$, $k^+=1$,
        				$K\approx85000 - 422000$. Data are averaged over 9 runs.}
			\label{fig:b_ana_I=100}
		\end{figure}	

			The most interesting behavior
			of the system is revealed when investigating the dependence on $b$.
			Figure \ref{fig:b_ana_I=100} shows this dependence for two different
			values of $I$.
			For a small value of $I$, $\rho_{init}$ rises 
			proportionally to	$b$.
			Increasing $I$ reveals, that, for a certain  value of
			$b$, $\rho_{init}$ reaches a maximum.
			The value of $b$ corresponds to the slope in the log-log plot of the
			size distribution.
			For a larger $b$ the probability to find a small
      particle will increase.
      The summed up length of the $K$ particles in the 
			system must be equal to the length $l$ of the interval for all values of
			$b$.
			Therefore, a system with a larger $b$ has more particles but
			their average size is smaller.

 			As mentioned in the previous section, the hierarchical initialization
      generates small holes. Thus, increasing $b$ increases
      $\rho_{init}$ because more smaller particles, able to fit into these
      holes, become available. 
      On the other hand, during each initialization, holes, smaller
			than $r_{min}$, are generated.  They will never be filled and
			increasing the number of particles $K$ in the system enhances the
			generation of theses holes.
			These two effects act against each other 
      and so there exists a value of $b$ for which the
      the initialization density $\rho_{init}$ will have a 
      maximum.

    \section{Conclusion}
  	We generalized the reversible Parking Lot Model introduced
  	by Krapivsky and Ben-Naim \cite{JChemPhys_100_9_6778} 
  	from a monodisperse model to a polydisperse one.
  	This 'self consistent reversible polydisperse Parking Lot Model' results
		in a final and stable state of very high density.
		
		The crucial part, in
		order to reach high densities, is the self consistent desorption probability.  The use
		of a constant desorption rate results in an exclusion of larger
		particles. The system will reach a final state of unrealistic low
		density. Thus, the small particles must not be allowed to block a larger
		interval and large particles already adsorbed in the interval need to
		stay.	This is realized through a hierarchical initialization where each particle
    gets $I$ trials for adsorption.
    		
    To keep the density on a high level we propose a self consistent desorption
		probability. A particle can only desorb if at least one
    hole, as large as this particle, exists.
		The density rises as a power law with the number of trials $I$ for
		each particle during the initialization and is fitted well by Eq.
		(\ref{eq:fit2}).
		The smaller the range $\overline{R}$, the lower the final density.
		
		The most interesting result is: There exists a value for the exponent $b$ 
		of the power law of the $t_0$ distribution
		for which the final density reaches a maximum as seen in Figure \ref{fig:b_ana_I=100}.

		\noindent
    In short: A large number of trials $I$ and a wide range of
    particles sizes $\overline{R}$ results in a high final density. 
    The maximal possible density is determined by the exponent
		$b$.
		
		In this paper we extended the work of Brilliantov \textit{et al.}
		\cite{PhysRevLett_76_21_4058,PhysRev_E_58_3_3530} by adding the self
		consistent desorption to the polydisperse RSA in one dimension. We could
		also apply the expression for the	evolution of density, measured
		experimentally by Knight \cite{Physrev_E_51_5_3957} and Nowak
		\cite{Physrev_E_57_2_1971}, to our density evolution.  As the
		initialization is the crucial point in our model it would be interesting
		to see the density evolution of an experiment using our hierarchical
		initialization.

		Future work should focus on the behavior of this model in two or three
		dimensions and under the influence of gravity. The main difficulty will
		be the development of a fast algorithm able to determine the
		hole distribution. 

\bibliographystyle{apsrev}
\bibliography{bib/sonstige,bib/vibgran.bib,bib/tetris,bib/sfb,bib/PLM,bib/apollonian,bib/buecher,bib/physrev_e,bib/powder_tech,bib/physrevlett}
\end{document}